\documentclass[aps,superscriptaddress,twocolumn,showpacs,showkeys]{revtex4}
\usepackage{graphicx}
\usepackage{amsmath}

\begin{document}

\title{Bonabeau model on a fully connected graph}

\author{K. Malarz}
\homepage{http://home.agh.edu.pl/malarz/}
\email{malarz@agh.edu.pl}
\affiliation{
Faculty of Physics and Applied Computer Science,
AGH University of Science and Technology,\\
al. Mickiewicza 30, PL-30059 Krak\'ow, Euroland
}

\author{D. Stauffer}
\affiliation{
Institute of Theoretical Physics,
Cologne University,
Z\"ulpicher Str. 77, D-50937 K\"oln, Euroland
}

\author{K. Ku{\l}akowski}
\affiliation{
Faculty of Physics and Applied Computer Science,
AGH University of Science and Technology,\\
al. Mickiewicza 30, PL-30059 Krak\'ow, Euroland
}

\date{\today}

\begin{abstract}
Numerical simulations are reported on the Bonabeau model on a fully connected graph,
where spatial degrees of freedom are absent.
The control parameter is the memory factor $f$. The phase transition is observed at 
the dispersion of the agents power $h_i$. The critical value $f_C$ shows a hysteretic 
behavior with respect to the initial distribution of $h_i$. $f_C$ decreases with the system
size; this decrease can be compensated by a greater number of fights between a global
reduction of the distribution width of $h_i$. The latter step is equivalent to a partial 
forgetting.
\end{abstract}



\maketitle

\section{Introduction}

The game theory is considered to be mathematical formulation of theory of conflicts
\cite{straf}. Since its initialization by von Neumann and Morgenstern in 1946 \cite{neumrg},
it becomes a powerful branch of knowledge, with numerous links to other areas of science.
Offering new insight into old problems, these links allow to widen our 
perspective to new interdisciplinary applications.

Such is the concept of self-organizing hierarchies in animal societies, introduced by Bonabeau {\it et al.} in 1995 \cite{bon} (see Refs. \cite{latest,sou,mart} for the most recent results and Ref. \cite{rev} for reviews).
In the Bonabeau model, $N$ agents walk randomly
on a given area and fight when they met. Initially, the outcome of fights is random.
However, agents are able to remember for some time their past results and this memory influences
their subsequent fights. Then, there are two competitive mechanisms. First, each fight 
influences the agents' power: winner is stronger and looser is weaker, what alters the 
probabilities of winning of their future fights. Second, the information of these alterations 
is gradually erased. As a consequence, a phase transition can be observed: for a given 
velocity of forgetting, frequent fights produce a hierarchy of permanent winners and 
permanent loosers. This hierarchy is maintained in time. However, if fights are rare,
the hierarchy is being forgotten quicker, than it is reproduced. The frequency of fights 
depends on the number of fighters on a given area. The order parameter is the dispersion
of power of the agents, or the dispersion of probabilities of winning/loosing of pairs
of agents. The phase transition was termed in \cite{sou} as the one between hierarchical 
and egalitarian society. Besides its sociological associations, the Bonabeau model offers
nontrivial dynamics of a specific, time-dependent spontaneous symmetry breaking, which 
--- in our opinion --- deserves attention from a theoretical point of view. Unfortunately, 
numerical experiments described in Ref. \cite{sou} have shown that the order and even the 
appearance of the phase transition depends on a direct way of forgetting, i.e. the time 
dependence of the agents' power, and on the presence of mean-field-like coupling which is 
absent in Ref. \cite{bon}.

A reasonable strategy of resolving this puzzle seems to separate it to elements as simple as 
possible, and to observe their properties. Here we propose a formulation where the spatial
coordinates of agents are absent. Instead, the agents are placed at nodes of a fully 
connected graph, i.e. each agent can meet every other agent. Scale free networks were mentioned in 
\cite{mart}. In next section the applied 
procedure is described in detail. Numerical results are reported and concluded in two 
subsequent sections.

\section{The model}

Two fighters $i$ and $j$ are selected randomly from a population of $N$ 
agents. The probability that $i$-th agent wins over $j$-th is

\begin{equation}
P(i,j,t)=\frac{1}{ 1+\exp\{\sigma(t)[h_j(t)-h_i(t)]\} }
\end{equation}
where $h_i$ is the power of $i$-th agent at time $t$ and 
\begin{equation}
\sigma^2(t)=\langle h_i^2(t)\rangle -\langle h_i(t)\rangle ^2,
\end{equation}
where $\langle\cdots\rangle $ is the average over $N$ agents. As an output of the fight,
the power $h_i$ 
of the winner increases by $\varepsilon$ and the one of the looser decreases by $\varepsilon$. 

Every $N_f$ steps (i.e. fights), the 
powers $h_i$ of all agents are multiplied by the factor $(1-f)$, where $f\in(0,1)$. This is the 
step of `forgetting'. 
As often as forgetting procedure takes place the current value of dispersion $\sigma$ is 
evaluated which is then fixed during next $N_f$ fights. The number of such updates of $\sigma$ 
is $N_{iter}$. Then, the total number of fights during one simulation is $N_fN_{iter}$.

\section{Results}

The parameters of the calculations are: the system size $N$, the number of fights $N_f$ between
the subsequent updates of $\sigma$, the number of steps $N_{iter}$ and the change $\varepsilon$ of 
the power $h_i$. The initial 
distribution of $h_i$ appears also to be relevant. This is set either random ($h_i\in[-N/2,N/2]$), or homogeneous, i.e. 
$h_i=i\varepsilon$ for all $i$, or delta-like, i.e. $h_i=0$ for all $i$. We keep $\varepsilon=0.01$. The 
output of the simulation is the critical value of $f$, i.e. $f_C$, where $\sigma$ changes abruptly.
We can speak about `hierarchical' (large $\sigma$) or `egalitarian' (small $\sigma$) society.
As a rule, for small $f$ we get hierarchy, and for large $f$ --- equality. 

\begin{figure}
\begin{center}
\includegraphics[scale=0.6]{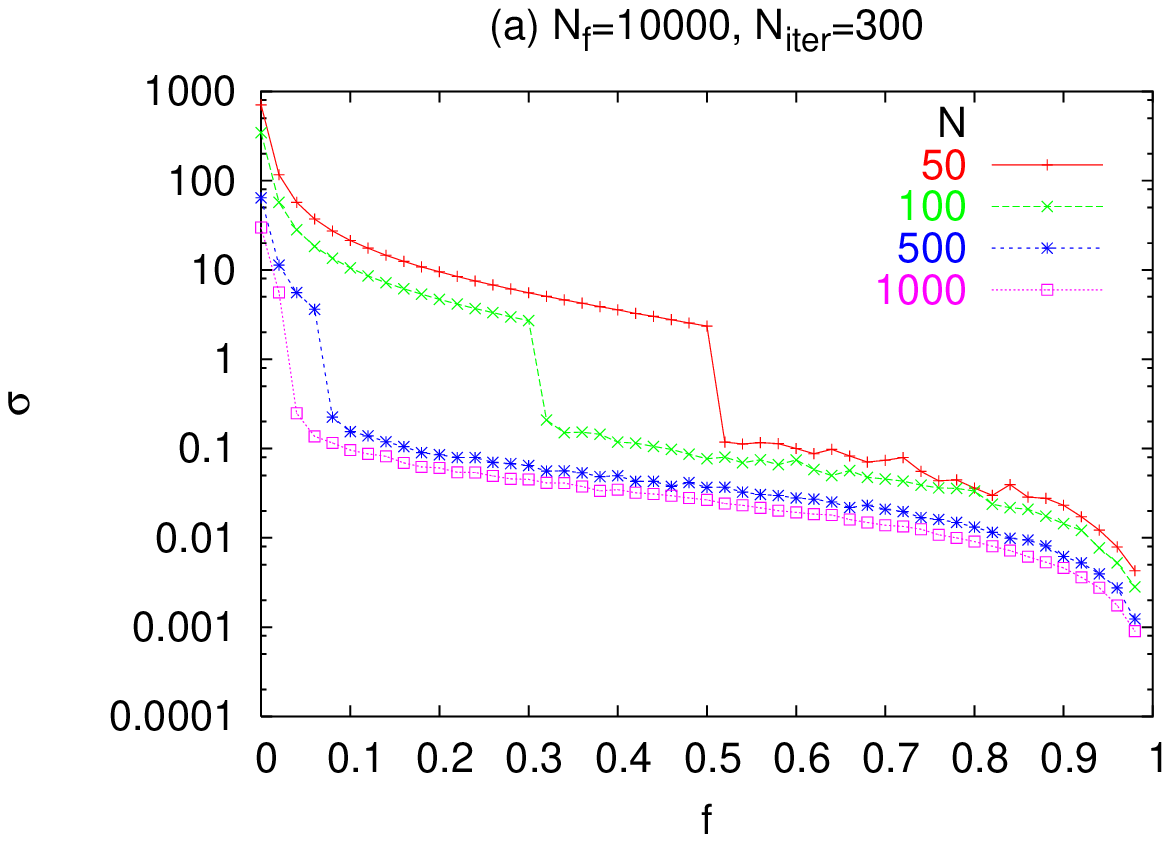}
\includegraphics[scale=0.6]{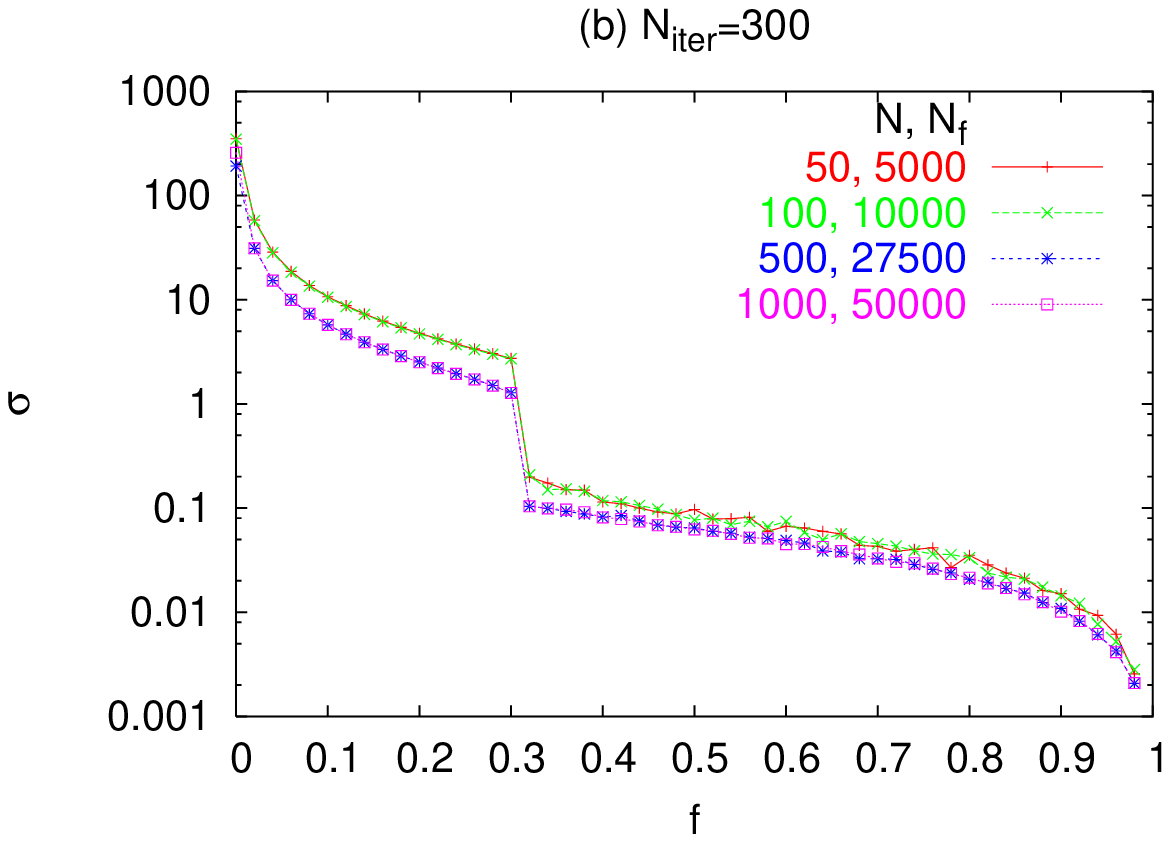}
\includegraphics[scale=0.6]{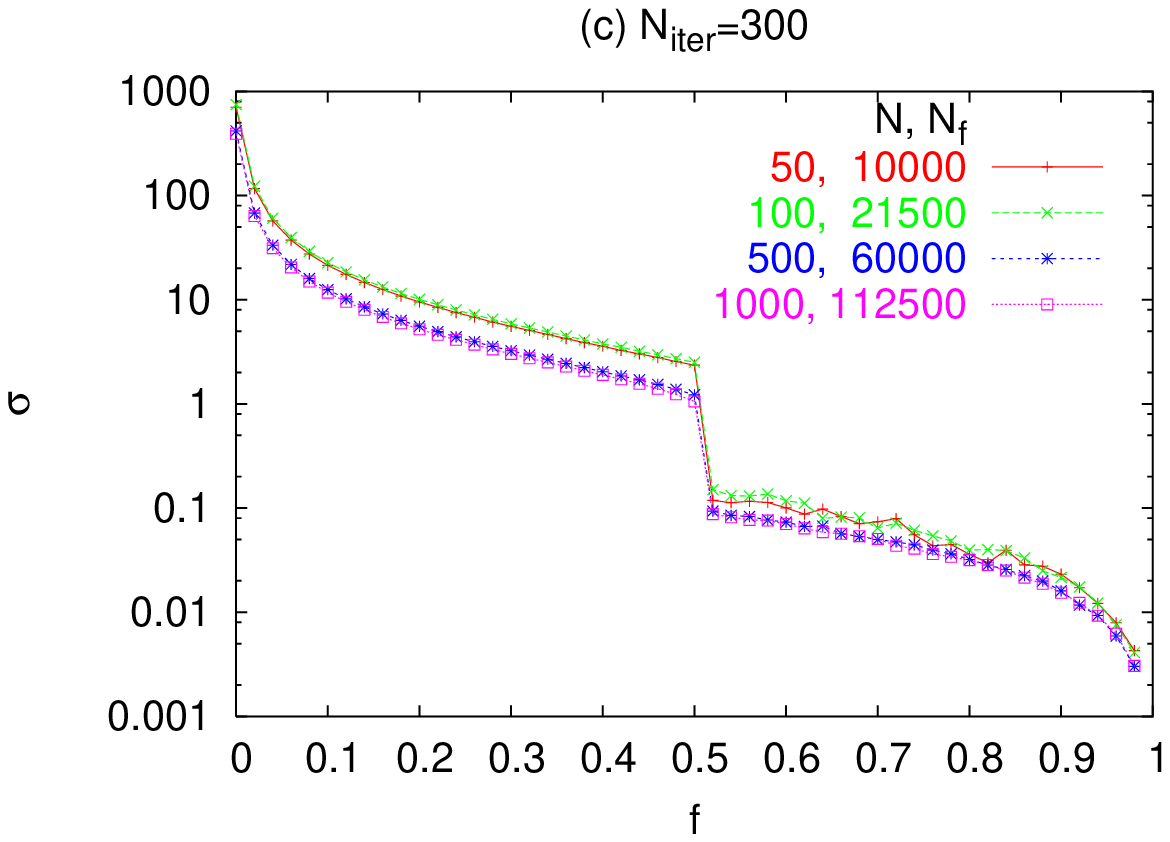}
\end{center}
\caption{Dependence of $\sigma(f)$ for (a) fixed $N_f=10^4$ and different values of $N$ and with tuned values of $N_f$ for $N$-independent values of $f_C$: (b) $f_C=0.3$, (c) $f_C=0.5$.}
\label{fig-1}
\end{figure}

\begin{figure}
\begin{center}
\includegraphics[scale=0.6]{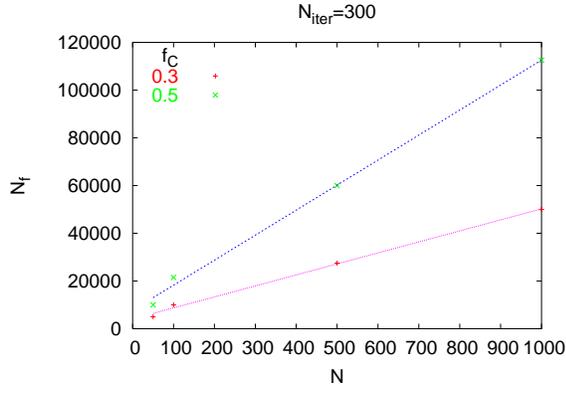}
\end{center}
\caption{Dependence of $N_f(N)$ for different values of $f_C$.
Linear fits are $N_f=105N+7756$ ($f_C=0.5$) and $N_f=46N+4095$ ($f_C=0.3$).}
\label{fig-2}
\end{figure}

It appears that $f_C$ depends on the ratio between $N$ and $N_f$. Keeping $N_f$ constant and
increasing $N$, we make forgetting more and more relevant, because each agent fights less
between subsequent forgettings. Then, the critical value $f_C$ decreases with $N$, as shown 
in Fig. \ref{fig-1}(a). We can compensate this variation, changing $N_f$ and $N$ simultaneously as to keep
$f_C$ constant. Two series of this procedure are shown in Figs. \ref{fig-1}(b) and \ref{fig-1}(c), for $f_C=0.3$ and $f_C=0.5$. In Fig. \ref{fig-2}, we show two respective curves of $N_f$ against $N$. 

\begin{figure}
\begin{center}
\includegraphics[scale=0.6]{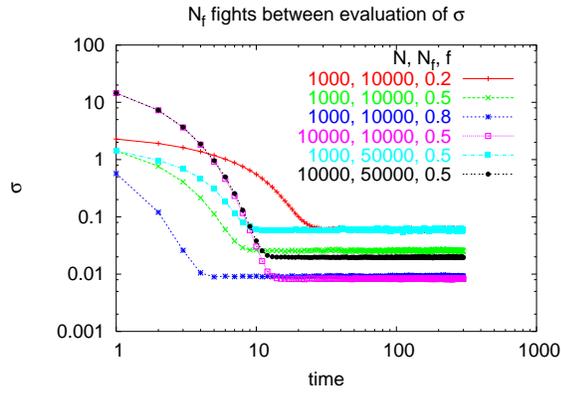}
\end{center}
\caption{Time evolution of $\sigma$.
Between subsequent $\sigma$ evaluations $N_f$ fights take place.}
\label{fig-3}
\end{figure}

We note that the obtained values of $\sigma$ are stable vs time, because the distribution 
of the agent's power $h_i$ stabilize after some transient time. Examples of this dynamics
are shown in Fig. \ref{fig-3} for various values of the parameters. 

\begin{figure}
\begin{center}
\includegraphics[scale=0.6]{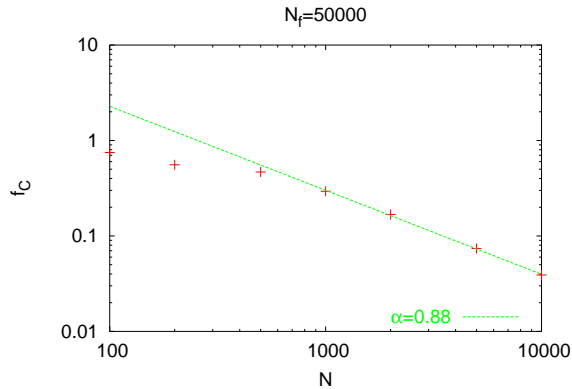}
\end{center}
\caption{Dependence of the critical value of the  forgetting parameter $f_C$ on the system size $N$ for $N_f=50000$.
The least-square fit $f_C(N)\propto N^{-0.88}$ to the last decade is included.}
\label{fig-4}
\end{figure}

According to what was said above, the system size of the critical value of 
forgetting parameter $f_C(N)$ decreases to zero for large $N$. The character of this variation
is shown in Fig. \ref{fig-4}. It is likely that there is a power-like behavior, i.e. $f_C\propto N^{-\alpha}$, with $\alpha\approx 0.88$.

\begin{figure}
\begin{center}
\includegraphics[scale=0.6]{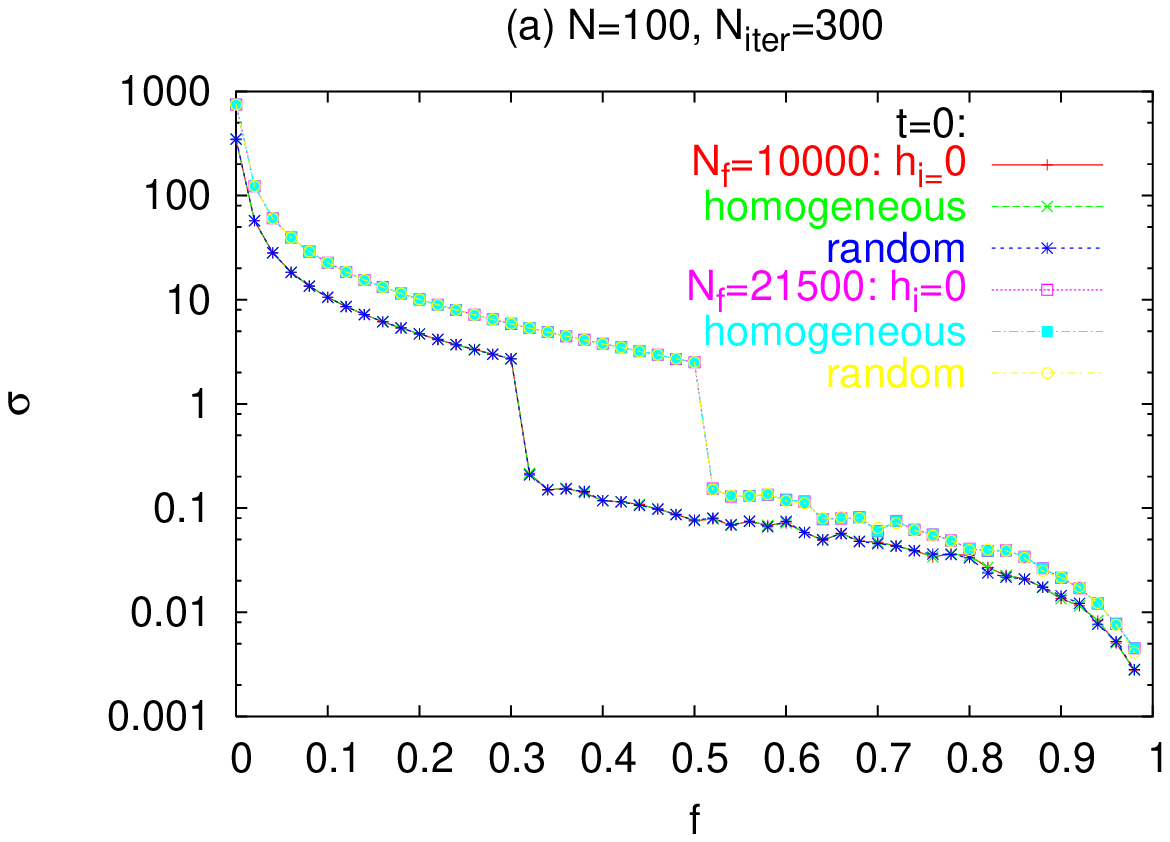}
\includegraphics[scale=0.6]{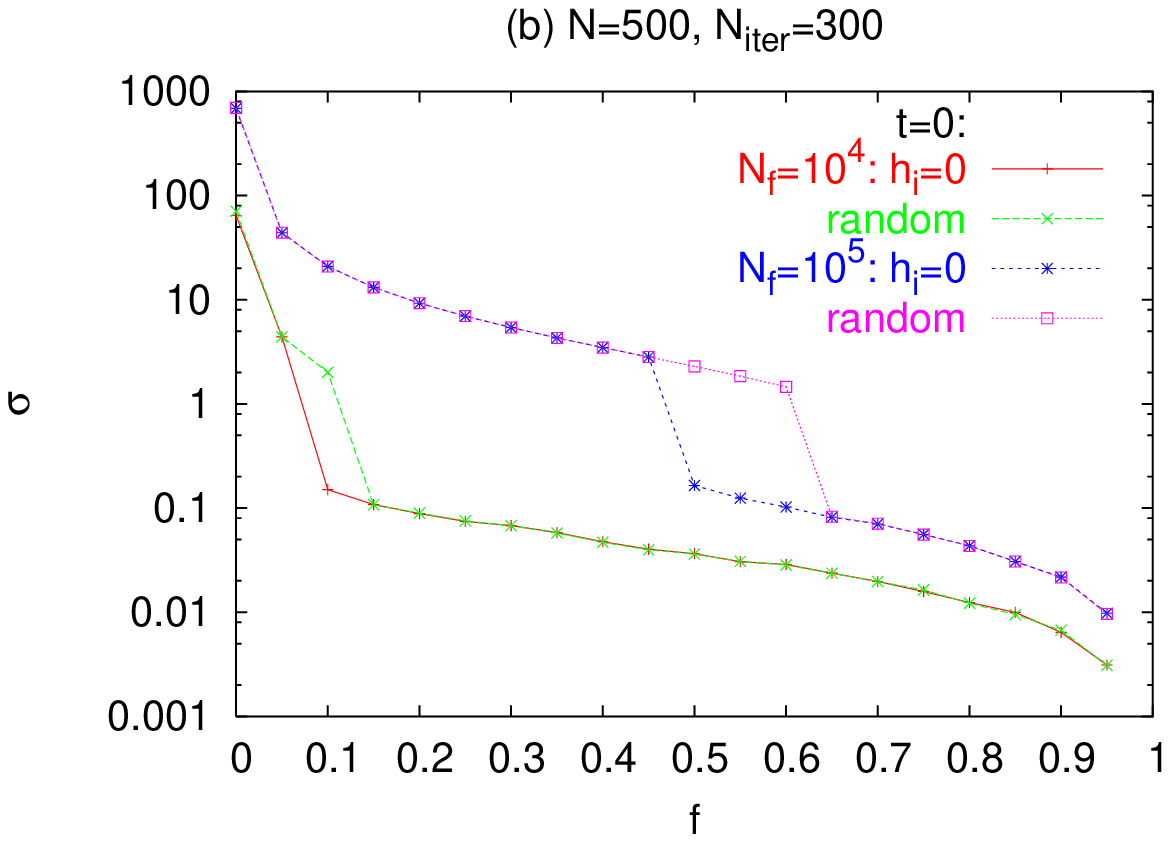}
\includegraphics[scale=0.6]{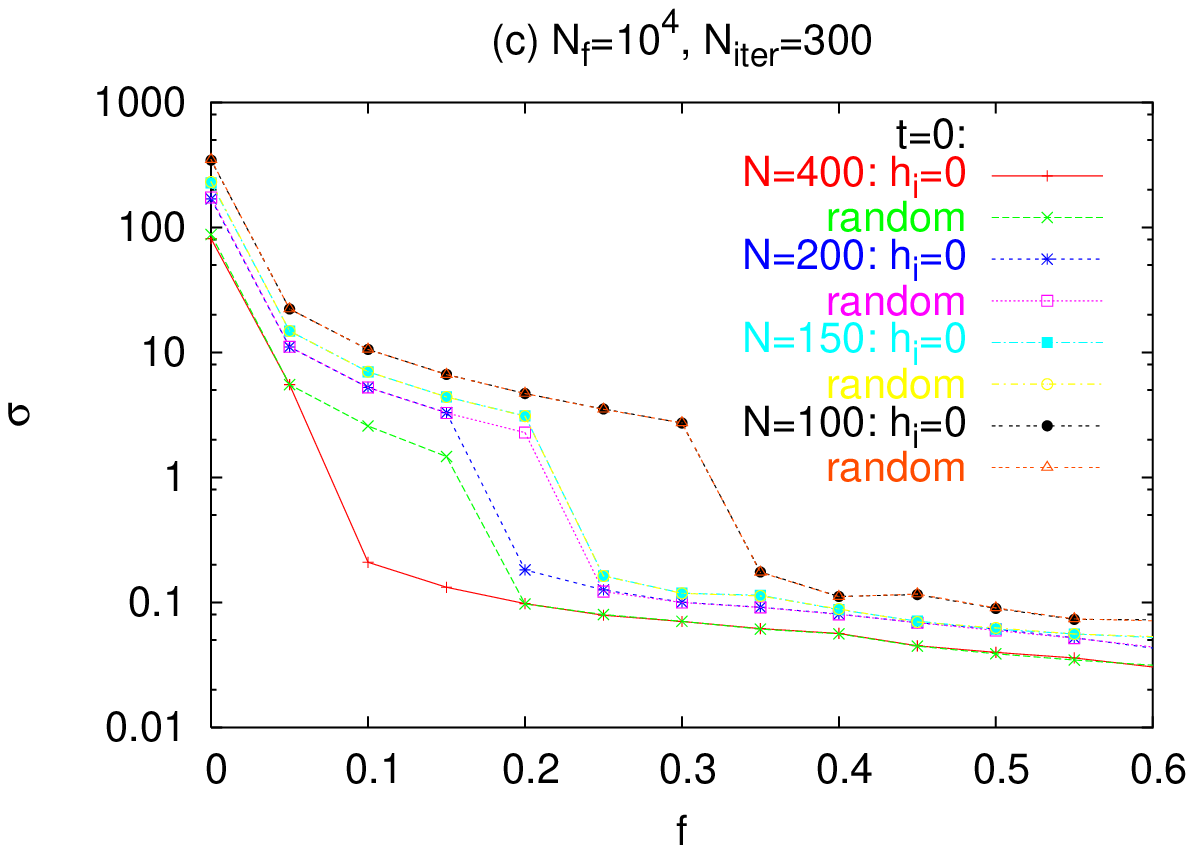}
\end{center}
\caption{
(a) Final values of $\sigma$ in small societies ($N=100$) do not depend 
on an initial distribution of $h_i$.
Different values of $N_f$ do not alter this result but shift the critical value $f_C$.
(b) When system is large enough ($N=500$), the critical point $f_C$ for given number of fights 
$N_f$ depends on the initial conditions.
(c) The initial inequality influences the critical point for $N\ge 175$.
}
\label{fig-5}
\end{figure}

For small systems, the final value of $\sigma$ does not depend on the initial distribution 
of $h_i$. However, above $N=175$ the final distribution of $h_i$ shows some hysteretic behavior,
i.e. $\sigma$ does depend on initial values of $h_i$. In Fig. \ref{fig-5} we show the curves $\sigma (f)$
obtained for random, homogeneous or delta-like initial values of $h_i$. The curves, identical
for $N<175$, split abruptly for larger systems. On the contrary, we checked that the 
hysteretic effect is not observed in the approach in Ref. \cite{sou}, at least near the critical 
point reported there.

\section{Discussion}

In their seminal paper \cite{bon}, the authors put several examples and suggestions on 
societies, which could self-organize into the hierarchical state. There, reference are
given on animal societies, like bees, wasps, ants, chickens, cows and ponies. A question 
arises how important are spatial degrees of freedom in these societies. This is a central 
task for our considerations, because the territorial aspect of the problem of hierarchy 
is disregarded here, while it was included in earlier work.
It seems that in small groups of some animals, the difference in 
hierarchy is not to be here or there but rather to do this or that. This is particularly
true in societies of some primates, as {\it Pan troglodytes}, {\it Homo sapiens} or 
{\it Gorilla gorilla} \cite{lorentz,morris,goodall}.

The Bonabeau model, once released, lives with its own life as it provides nontrivial
questions on its mathematical properties. However, its core is the sociological application:
description of a group of agents, possibly human, which concentrate on their hierarchy.
Such hierarchy appears in a natural way in sports competitions.
For example, each football league in the world produces one table at the end of the season
with the numbers of points and goals for each team.
Usually, the extremes at the top and the bottom have widely spaces points, while in the center
the average teams differ just by one point.
This is not dissimilar to the power distribution of the Bonabeau model \cite{mart}.

How important is the hierarchy for the agents, depends on their experience and social environment. 
Except army and some universities, we are interested in formation of a society rather 
egalitarian than hierarchical, and tasks of educational organizations following this attitude 
are well known \cite{newc}. In the Bonabeau model, these efforts get a well-defined purpose: 
to be on the right side of the transition.

\bigskip

\noindent
{\bf Acknowledgments.}
The numerical calculations were carried out in ACK\---CY\-F\-RO\-NET\---AGH.
The machine time on HP Integrity Superdome is financed by the Polish Ministry of Science and Information Technology under grant No. MNiI/\-HP\_I\_SD/\-AGH/047/\-2004.



\begin{thebibliography}{88}
\bibitem{straf} Ph. D. Straffin, {\it Game Theory and Strategy}, Math. Association of America, Washington, D. C. 1993.

\bibitem{neumrg} J. von Neumann, O. Morgenstern, {\it Theory of Games and Economic Behavior}, Wiley 1967 (first ed. 1944).

\bibitem{bon} E. Bonabeau, G. Theraulaz, J.-L. Deneubourg, Physica {\bf A217}, 373 (1995).

\bibitem{latest}
G. G. Naumis, M. del Castillo-Mussot, L. A. Perez, G. J. Vazquez,
to appear in Int. J. Mod. Phys. {\bf C17}, (Jan 2006);
L. K. Gallos,
Int. J. Mod. Phys. {\bf C16}, 1329 (2005);
C. Schulze, D. Stauffer,
Adv. Complex Syst. {\bf 7}, 289 (2004);
A. O. Sousa, D. Stauffer,
Int. J. Mod. Phys.  {\bf C11}, 1063 (2000).

\bibitem{sou} D. Stauffer,
Int. J. Mod. Phys.  {\bf C14}, 237 (2003). 

\bibitem{mart}
D. Stauffer, J. S. S\'a Martins,
Adv. Complex Syst. {\bf 6}, 559 (2003).

\bibitem{rev}
D. Stauffer,
{\tt physics/0503128};
D. Stauffer,
Comput. Sci. Eng. {\bf 5}, 71 (2003);
D. Stauffer,
Fractals {\bf 11}, 313 (2003).

\bibitem{lorentz} K. Lorentz, {\it On Aggression}, Harcourt, Orlando 1966.

\bibitem{morris} D. Morris, {\it Naked Ape}, McGraw Hill, New York 1967.

\bibitem{goodall} J. Goodall, {\it In the Shadow of Man}, Houghton Mifflin Comp., Boston 1988.

\bibitem{newc} Th. M. Newcomb, R. H. Turner, Ph. E. Converse, {\it Social Psychology}, Rinehart
and Winston, Inc., New York 1965, Chapt. 15.

\end{thebibliography}
\end{document}